# Intrinsically patterned two-dimensional materials for selective adsorption of molecules and nanoclusters

X. Lin[1,†], J. C. Lu[1,†], Y. Shao[1,†], Y. Y. Zhang[1,2,†], X. Wu[1], J. B. Pan[1], L. Gao[1], S. Y. Zhu[1], K. Qian[1], Y. F. Zhang[1], D. L. Bao[1], L. F. Li[1], Y. Q. Wang[1], Z. L. Liu[1], J. T. Sun[1], T. Lei[3], C. Liu[3], J. O. Wang[3], K. Ibrahim[3], D. N. Leonard[4], W. Zhou[1,4], H. M. Guo[1], Y. L. Wang[1,*], S. X. Du[1,*], S. T. Pantelides[2,1], H.-J. Gao[1,*]

1. Institute of Physics & University of Chinese Academy of Sciences, Chinese Academy of Sciences, Beijing 100190, P. R. China

2. Department of Physics and Astronomy and Department of Electrical Engineering and Computer Science, Vanderbilt University, Nashville, Tennessee 37235, USA

3. Institute of High Energy Physics, Chinese Academy of Sciences, Beijing 100049, P. R. China

4. Material Sciences and Technology Division, Oak Ridge National Laboratory, Oak Ridge, Tennessee 37381, USA

†These authors contributed equally to this work and are listed alphabetically.

*Corresponding author. E-mail: ylwang@iphy.ac.cn (Y.L.W); sxdu@iphy.ac.cn (S.X.D); hjgao@iphy.ac.cn (H.J.G)

**Two-dimensional (2D) materials have been studied extensively as monolayers [1-5], vertical or lateral heterostructures [6-8]. To achieve functionalization, monolayers are often patterned using soft lithography and selectively decorated with molecules [9,10]. Here we demonstrate the growth of a family of 2D materials that are intrinsically patterned. We demonstrate that a monolayer of $PtSe_2$ can be grown on a Pt substrate in the form of a triangular pattern of alternating 1T and 1H phases. Moreover, we show that, in a monolayer of CuSe grown on a Cu substrate, strain relaxation leads to periodic patterns of triangular nanopores with uniform size. Adsorption of different species at preferred pattern sites is also achieved, demonstrating that these materials can serve as templates for selective self-assembly of molecules or nanoclusters, as well as for the functionalization of the same substrate with two different species.**

Two-dimensional (2D) materials have unique and unusual properties that promise novel applications [6,11,12]. Typically, 2D materials need to be functionalized and patterned, just as semiconductor films need to be doped *n*-type or *p*-type in a lithographically-patterned way to produce *pn* junctions (light-emitting diodes, junction lasers, resistors) and transistor structures [10,11]. Functionalization of 2D materials is done in a variety of ways, e.g., by doping with impurities or molecules, either within the 2D lattices or adsorbed on the exposed surfaces [10]. In particular, functionalization by adsorbates can facilitate catalysis, sensing, optical or magnetic response, nanodevices, or other applications. To achieve selective or multiple functionalization by adsorbates, patterning by soft lithography has been used [9]. 2D materials exhibiting forms of atomic-scale patterning have been proposed based on calculations by the choice of Bravais lattice [13] or stoichiometry [14,15]. Moiré patterns represent a nano-scale form of patterning imposed by the substrate on otherwise homogeneous 2D materials [16,17].

Herein, we report the fabrication of prototypes of a family of 2D materials that are intrinsically patterned on the nano-scale and can be selectively or dually functionalized by adsorbates. We report two prototype 2D materials, each obtained by direct selenization of a monatomic metal substrate. More specifically, we show that, by controlling the annealing temperature and deposition amount of Se on a Pt substrate, we obtain either a homogeneous monolayer $PtSe_2$ in the 1T phase or a "patterned monolayer" comprising a periodic triangular structure of alternating 1H and 1T phases. The size of the triangles can be tuned by varying the density of deposited Se atoms. The choice between a homogeneous 1T phase or patterned 1H/1T phases is reversible. We also report the fabrication of a patterned monolayer of CuSe having periodic arrays of triangular nanopores, with distinct domains. In both patterned materials, adsorption of different species at preferred pattern components has been achieved, illustrating the potential for selective/dual functionalization. In the above two prototype intrinsically-patterned monolayers, we show that $Fe_{13}Se_7$ nanoclusters, a known catalyst [18], can be grown selectively in the nanopores of patterned CuSe monolayer, providing a way to avoid aggregation in nanoparticle catalysts. Another molecular species, iron(II) phthalocyanine (FePc), can be selectively adsorbed on the CuSe surface between the nanopores. Throughout the exposition we use density-functional-theory (DFT) calculations to account for the observations.

A prototype of a 2D intrinsically-patterned material is shown schematically in Fig. 1a. The entire monolayer is PtSe$_2$, but the two distinct types of triangles are two different phases, 1H and 1T (blue and yellow triangles, respectively). A large-scale STM image of such a structure is shown in Fig. 1b. We will describe shortly how the patterned structure was fabricated and how we determined that the two kinds of moiré patterns, exhibiting honeycomb and hexagonal features, are the 1H and 1T phase of PtSe$_2$, respectively. The honeycomb and hexagonal structures are more visible in the intermediate-resolution STM images of Fig. 1c to 1e. These images also demonstrate that triangles of different sizes are possible. Atomic-resolution images are shown in Fig. 1f to 1h, together with simulations in Fig. 1i to 1k. The detail structures are shown in Supplementary Information Fig. S1 to S3. A clean, sharp interface is clearly visible in Fig. 1g.

The formation of a homogeneous monolayer 1T-PtSe$_2$ on a Pt(111) substrate by direct selenization has been described in Ref. 19. Such a monolayer, fabricated by annealing at 270°C during the selenization of a Pt substrate is shown in Fig. 2a. A homogeneous 1T sample transforms to a triangular pattern (Fig. 2b) by annealing at 400°C. This elevated temperature leads to the loss of some selenium, leaving vacancies (marked by the yellow arrow) in the film. The transformation process is monitored by the change of chemical states of Se atoms via in-situ X-ray photoelectron spectroscopy (XPS) measurements. As shown in Fig. 2c, the Se 3$d$ core level from the pure 1T structure PtSe$_2$ monolayer exhibits two peaks (colored in red). After annealing the sample to 400°C, the intensity of the initial red peaks decreases and two new peaks (colored in blue) appear. The new peaks have an energy shift of 0.4-eV relative to the initial ones. Considering that a similar energy shift in the XPS spectra has been reported between H- and T-phases in MoTe$_2$ flakes [11], the XPS energy shift (Fig. 2c) in PtSe$_2$ is consistent with a local 1T-to-1H transformation. Theoretical calculations indicate that free-standing 1H-PtSe$_2$ monolayer is not stable (Supplementary Information Fig. S4). For the patterned structure with alternating triangular 1T and 1H areas, the 1H domains are limited in size and are likely stabilized by the surrounding 1T-PtSe$_2$.

We also measured I(z) spectra, which provide information on the local work function. As shown in Supplementary Information Fig. S5, the derived work function in 1T domains is larger than that in 1H domains, which is confirmed by DFT calculations. These results provide further evidence of the existence of 1H-PtSe$_2$.

The transformation of a pure 1T phase to a patterned 1H/1T phase is reversible. By adding Se atoms and annealing the sample to ~270°C, the triangular pattern reverts to a defect-free homogeneous 1T structure (Fig. 2a). The extra peaks in the XPS spectrum of the patterned 1H/1T phase (blue peaks in Fig. 2c) also disappear, confirming that the sample reverts to a pure 1T structure. The disappearance of the Se vacancies during the reversible transformation provides evidence that Se vacancies mediate the formation of 1H domains.

Based on the above experimental data, we use the schematic drawings in Fig. 2d to g, to show the transformation process and mechanism. Specifically, thermal annealing of a pristine 1T structure PtSe$_2$ film at an elevated temperature leads to the loss of top-layer Se atoms (depicted by the arrows in Fig. 2b), generating Se vacancies (Fig. 2e). DFT calculations find that lines of vacancies are energetically preferred (see Supplementary Information Fig. S6) which is in agreement with a previous report [20]. Because of the threefold point symmetry, three directions of such lines are possible, which leads to the formation of triangles, as shown in Fig. 2f. Within the triangle surrounded by vacancies, a transformation of the 1T phase to the 1H phase occurs and leads to a 1H/1T patterned structure. A similar structural transformation has been reported in a MoS$_2$ monolayer [21]. More details of the 1T-to-1H transformation in PtSe$_2$ can be found in Supplementary Information Fig. S6 and S7.

An alternative method to produce the 1H/1T triangular pattern is to control the initial density of Se atoms. This process is shown schematically in Supplementary Information Fig. S8. Whereas for the fabrication of a homogenous 1T-PtSe$_2$, the crucible is filled with the Se source and held at-center for a long time to deliver ample amounts of Se everywhere on the sample. While the 1H/1T triangular pattern is fabricated by holding the crucible filled with the Se source off-centered to the substrate, resulting in a Se density gradient, which subsequently leads to different structures. A triangular tiling pattern of 1H/1T-PtSe$_2$ forms in the area with a relatively low Se density. As the density of Se atoms increases, the size of 1T-PtSe$_2$ triangles increases and the number of 1H-PtSe$_2$ triangles decreases significantly. As the Se density increases further, it eventually leads to a defect-free homogeneous 1T-PtSe$_2$ film. An intermediate structure can be found in between the 1H/1T-tiling region and the homogeneous 1T region. Typical STM images with different Se amounts are shown in Supplementary Information Fig. S8. The XPS mapping of Se 3$d$ electrons along the direction of increasing Se density areas (area ① to area ④) confirms a significant decrease of 1H-PtSe$_2$ and increase of 1T-PtSe$_2$. These data confirm the role of Se vacancies and suggest that a critical concentration is necessary to produce a periodic triangular pattern.

The highly-ordered triangular tiling patterns constructed by 1H and 1T phases of a monolayer PtSe$_2$ make it a suitable template for selective adsorption of different species. Figures 2h and 2i show a large-scale area and zoomed-in images of pentacene (C$_{22}$H$_{14}$) on patterned PtSe$_2$. It is clear that pentacene molecules adsorb selectively on 1H triangles, allowing the formation of a periodic array of molecules. The high-resolution STM image shown in Fig. 2i demonstrates more details for the nanoclusters, which present three orientations corresponding to the triangular shape of the 1H-PtSe$_2$ region.

The preference of molecules to bind on the 1T or 1H domain is determined by the relative binding energies. As shown in Table S1, all the tested molecules have larger binding energies on the 1H domain. The basic idea, nevertheless, is that different molecules can in principle adsorb on the two phases of an intrinsically-patterned 2D material, leading to simultaneous dual functionalization for catalysis or other applications. The possibility of different objects adsorbing on different regions of a patterned material will be demonstrated in the next prototype intrinsically-patterned 2D material, namely monolayer CuSe on a Cu substrate.

In addition to 1H/1T patterned PtSe$_2$, we have fabricated an intrinsically-patterned form of monolayer CuSe by direct selenization of Cu(111) substrate (more details of the fabrication process are given in Supplementary Information). A large-scale STM image of a monolayer CuSe with periodic arrays of identical triangular nanopores is shown in Fig. 3a. The nanopores form a hexagonal lattice as indicated by the green hexagonal lattice matrix with a periodicity of ~ 3 nm. The figure clearly shows domains with triangular nanopores of opposite orientations, separated by domain boundaries comprising parallelogram-shaped nanopores. High-resolution STM images of a domain-boundary region and a single nanopore are shown in Fig. 3b and 3c, respectively. The shape of the nanopores, honeycomb lattice of CuSe, and the zigzag edges of the nanopore are clearly resolved.

An atomic-resolution STM image of a patterned monolayer CuSe with nanopores is shown in Fig. 3d. Based on a low-energy-electron-diffraction (LEED) pattern (as shown in Supplementary Information Fig. S9), a ($4\sqrt{3}\times4\sqrt{3}$) CuSe sitting on a ($11\times11$) Cu(111) model is proposed. The optimized atomic configuration is shown in Fig. 3f. For better visualization, the Cu(111) substrate is hidden. The simulated STM image based on the proposed atomic model is shown in Fig. 3e, in excellent agreement with the experimental data. Note that the wiggly atomic lines (marked by purple curves in Fig. 3d to f) induced by the release of stress are well reproduced by the DFT simulations.

To further confirm that the fabricated CuSe is a monolayer, we performed a cross-section high-angle annular-dark-field (HAADF) scanning transmission electron microscopy (STEM) study. An experimental image, a simulated image and side view of the structure are shown in Fig. 3, g to i, respectively. The excellent agreement further proves that the CuSe pattern with periodic nanopores is a monolayer material.

The periodicity and size of the nanopores can be explained as follows. In principle, the Cu substrate and a CuSe monolayer are incommensurate. The two lattices can be made commensurate, however, if a ($4\sqrt{3}\times4\sqrt{3}$) CuSe supercell is stretched by 1.5%. This supercell then registers on a ($11\times11$) Cu(111) surface supercell, with a central Cu atom of the monolayer supercell on top of a Cu atom on the substrate. Removal of this central atom and three shells of atoms around it, a total of 13 atoms, produces the observed nanopores at the observed periodicity. In Supplementary Information Fig. S10, we demonstrate that removal of these 13 atoms amounts to removing three complete hexagonal rings and corresponds to maximum gain of energy (negative formation energy) compared with other options for removing atoms. We conclude that formation of 13-atom nanopores at the observed periodicity is the energetically preferred way to relieve the lattice-mismatch strain.

We have been able to demonstrate selective adsorption on the patterned CuSe monolayer with nanopores. Fe atoms deposited on a CuSe monolayer form W-shape clusters in the nanopore hexagonal superstructure as shown in the large-scale STM image in Fig. 4a. Considering the threefold symmetry of a nanopore, three different orientations of W-shape clusters can be found, as indicated by three "W" letters in Fig. 4a. DFT calculations suggest that the W-shape cluster contains 13 Fe atoms (the atomic configuration and a discussion about other possibilities are presented in Supplementary Information Fig. S11). Simulation of the STM image based on this model is provided in Fig. 4c, showing excellent agreement with the atomic-resolution STM image (Fig. 4b). The proposed model is further confirmed by *in-situ* XPS measurements. As shown in Fig. 4d, the Fe $2p_{3/2}$ and Fe $2p_{1/2}$ core-level spectra are located at 707.24 eV and 720.40 eV, respectively, which are similar to the Fe $2p$ core levels in bulk FeSe [22]. In addition to the W-shape clusters in the periodic nanopores, zigzag-shaped clusters also form in the more extended parallelogram-shaped nanopores at domain boundaries (see Supplementary Information Fig. S12).

In addition to the selective adsorption in nanopores, additional selective functionalization is achieved by deposition FePc molecules on the terraces. An atomic model of FePc is shown in the inset of Fig. 4f. As shown in Fig. 4e, when we deposit FePc molecules onto a patterned monolayer CuSe, the molecules are monodispersed and adsorbed only on CuSe terraces. A zoomed-in STM image (Fig. 4f) shows a clear cross feature, which is a typical apparent topography of FePc molecules [23]. It is noticed that FePc molecules on a CuSe substrate have preferred orientations. The angles between FePc molecules (indicated by the sketched purple cross) and the CuSe[$1\bar{1}0$] direction (white line) is ~45°. As a result of the threefold symmetry of the substrate and the fourfold symmetry of the molecule, there are three equivalent preferred orientations as indicated by the three purple crosses in Fig. 4e.

Comparing the two intrinsically-patterned monolayers, i.e. 1H/1T-PtSe$_2$ and CuSe with periodic nanopores, they are both achieved by direct selenization. After annealing at a high temperature, monolayer 1T-PtSe$_2$ undergoes a phase transition and forms the 1H/1T-pattern, while periodic nanopores appear in CuSe monolayer. We attribute the difference in the two prototype systems to the interactions between the epitaxial monolayer and the substrate. In the PtSe$_2$/Pt(111) system, the van der Waals interaction between PtSe$_2$ and the Pt substrate gives freedom to flip the structure from 1T to 1H. In contrast, the interaction between CuSe monolayer and a Cu substrate is much stronger. Emission of atoms and formation of nanopores becomes the way to lower the energy.

Both the PtSe$_2$ and CuSe samples are stable in ambient conditions. The samples were removed from the high-vacuum chamber and were kept in air without protection for more than 12 hours. The samples were then put back to the high-vacuum chamber and annealed at 200 °C for PtSe$_2$ and 400 °C for CuSe to remove the possible adsorbates. STM images show that the samples keep their patterns (see Supplementary Information Fig. S13). The stability in air makes these patterned monolayers proper for potential applications involving ambient conditions.

In summary, we demonstrate a concept of intrinsically-patterned 2D materials which can be selectively functionalized. We demonstrate the fabrication of patterned 2D monolayers in two cases: 1H/1T monolayer PtSe$_2$ tiling pattern and monolayer CuSe with periodic nanopores. Selective adsorptions are achieved in both cases. This work ushers a frontier for the fabrication of intrinsically patterned 2D materials.

**Acknowledgements**

We acknowledge the financial support from National Key Research & Development Projects of China (2016YFA0202300), the National Basic Research Program of China (2013CBA01600), the National Natural Science Foundation of China (Nos. 61390501, 51572290, 61306015 and 61471337, 51325204) and the Chinese Academy of Sciences (Nos. 1731300500015, XDB07030100, and the CAS Pioneer Hundred Talents Program). A portion of the research was performed in CAS Key Laboratory of Vacuum Physics. Work at Vanderbilt (STP and YYZ) was supported by the U.S. Department of Energy under grant

DE-FG02-09ER46554 and by the McMinn Endowment. Computations by YYZ were carried out at the National Energy Research Scientific Computing Center, a DOE Office of Science User Facility supported by the Office of Science of the U.S. Department of Energy under Contract No. DE-AC02-05CH11231. The electron microscopy work was supported in part by the U.S. Department of Energy, Office of Science, Basic Energy Science, Materials Sciences and Engineering Division, and through a user project at ORNL's Center for Nanophase Materials Sciences, which is a DOE Office of Science User Facility.

**Author contributions**

H.J.G. and S.T.P. conceived and coordinated the research project. X.L. designed the CuSe experiments. J.C.L. and K.Q. prepared CuSe samples and performed the STM experiments. Y.L.W. designed the $PtSe_2$ experiments. Y.S., X.W., S.Y.Z., L.F.L., Y.Q.W., Z.L.L., and H.M.G. prepared $PtSe_2$ samples and performed the STM experiments. T.L., C.L., J.O.W. and K.I. provided supports for XPS experiments. D.N.L. and W.Z. performed the STEM experiments. Y.Y.Z., J.B.P., L.G., Y.F.Z., D.L.B., and J.T.S performed the DFT calculations under the guidance of S.X.D. All authors participated in discussing the data and editing the manuscript.

**Additional information**

Supplementary information is available in the online version of the paper. Reprints and permissions information is available online at www.nature.com/reprints. Correspondence and requests for materials should be addressed to Y.L.W., S.X.D., or H.-J.G.

**Competing financial interests**

The authors declare no competing financial interests.

## Methods:

**Preparation of 1H/1T PtSe$_2$ on Pt(111) substrate by transforming from 1T PtSe$_2$.**
Single-layer 1T PtSe$_2$ films were grown in a commercial ultrahigh vacuum (UHV) system (Omicron), with a base pressure better than $3\times10^{-10}$ mbar, equipped with standard MBE capabilities. The Pt(111) substrate was cleaned by several cycles of argon-ion sputtering followed by annealing until sharp (1×1) diffraction spots in the LEED pattern and clean surface terraces in the STM images were obtained. High-purity Se (99.99%, Sigma-Aldrich) evaporated from a Knudsen cell was deposited onto the clean Pt(111) surface at room temperature. The sample was subsequently annealed up to 270°C to achieve selenization and crystallization, which leads to homogeneous 1T PtSe$_2$ films on Pt(111) substrate. By annealing this sample at 400°C, the 1T PtSe$_2$ transforms to the 1H/1T PtSe$_2$ with triangular patterns.

**Preparation of 1H/1T PtSe$_2$ on Pt(111) substrate by controlling the Se initial density.**
As shown in Supplementary Information Fig. S4, an alternative method to produce the 1H/1T triangular pattern is to control the Se initial density. Having obtained the clean Pt(111) surface, the crucible filled with the Se source was aligned at the edge instead of the center of the substrate during the fabrication process. This off-center alignment leads to varying initial density of Se atoms at different regions of the substrate, resulting in a Se density gradient. The sample was subsequently annealed up to 270°C to achieve selenization and crystallization, which finally leads to different structures. A triangular tiling pattern of 1H/1T PtSe$_2$ forms in the area with a relative low initial Se density. As the density of Se atoms increases, the size of 1T PtSe$_2$ triangles increases and the number of 1H PtSe$_2$ triangles decreases significantly. As the Se density increases further, it finally leads to a defect-free homogeneous 1T PtSe$_2$ film.

**Preparation of patterned CuSe, Fe/CuSe and FePc/CuSe on Cu(111) substrate.** The growth of 2D patterned CuSe was carried out in a commercial ultrahigh vacuum (UHV) system (Omicron) (base pressure better than $1\times10^{-10}$ mbar) equipped with a sample preparation chambers. Clean single-crystal Cu(111) (MaTecK) surface was obtained *via* cycles of argon-ion sputtering followed by annealing at 550°C. The cleanliness of the Cu(111) surface was verified by STM scanning. The 2D patterned CuSe sample was obtained by Se (99.99%, Sigma-Aldrich) deposition from a standard Knudsen cell while the Cu substrate was held at room temperature, and post-annealing at 200°C. For the Fe/CuSe and FePc/CuSe samples, the beam of Fe (99.99%, ESPI) and the FePc molecules (99.99%, Sigma-Aldrich) were generated from a commercially available e-beam cell

(Omicron) and a standard Knudsen cell, respectively. During the deposition, the patterned CuSe sample was kept at room temperature. After preparation, the samples were transferred to the Low-Temperature-STM head operating at ~ 4 K.

***Ex-situ* STEM characterization of CuSe on Cu(111) substrate.** *Cross-section TEM sample preparation.* A layer of amorphous carbon of ~100 nm in thickness was first deposited onto the as-grown CuSe monolayer on a Cu substrate shortly after removal from the MBE chamber. A TEM cross-section sample was prepared along the Cu[112] zone axis using focused ion beam (FIB) and was further thinned to ~40 nm thick using low-energy ion milling. *STEM-ADF (scanning transmission electron microscopy - annular dark field) imaging.* The experimental STEM-ADF imaging was performed on an aberration corrected Nion UltraSTEM-100 operated at 100 kV. The convergence semi-angle was set to 30 mrad, and the ADF images were collected from ~ 80-200 mrad semi-angle range. The thickness of the cross-section sample was measured to be ~ 40 nm using electron energy loss spectroscopy. STEM-ADF image simulation was performed using QSTEM with the same accelerating voltage, probe-forming angle and detector angle as those used in experiment. A single-crystal CuSe domain was used in the simulation, with the thickness of the cross-section sample set to 22 nm.

**STM and LEED measurements of the as-grown samples.** STM images of the samples were acquired in the constant-current mode, using an electrochemically etched tungsten tip. All voltages were applied to the sample with respect to the tip. The Nanotec Electronica WSxM software was used to process the STM images. Low-energy electron diffraction (LEED) was employed with a 4-grid detector (Omicron Spectra LEED) in a connected UHV chamber to identify the superstructure macroscopically.

**XPS measurements of the as-grown samples.** The *in situ* X-ray photoelectron spectroscopy (XPS) measurements of the samples were conducted at the Beijing Synchrotron Radiation Facility (BSRF). After growth, the samples were transferred by a UHV suitcase to the XPS station for measuring the change of chemical states of elements. The monochromatic synchrotron radiation is realized by four high-resolution gratings and is controlled by a hemispherical energy analyzer. The photon energy is in the range from 10 to 1100 eV.

**Theoretical calculations.** Quantum mechanical calculations based on density functional theory (DFT) are performed using the Vienna ab initio simulation package (VASP) [24,25]. The projector augmented wave method is employed to describe the core electrons. The local density approximation (LDA) [26,27] is used for exchange and correlation. The rotationally invariant LDA+U formalism proposed by Dudarev *et al.* is used [28] and $U_{eff}$ is chosen as 6.52 eV and 4.3 eV for Cu and Fe respectively [29]. The electronic wave functions are expanded in plane waves with a kinetic energy cutoff of 400 eV. The periodic slab models of PtSe$_2$/Pt(111) include four layers of (4×4) Pt(111) substrate, and monolayer of (3×3) PtSe$_2$. The k-points sampling is 5×5×1, generated automatically with the origin at the Γ-point. The periodic slab models of CuSe/Cu(111) include three layers of (11×11) Cu(111) and (4√3×4√3) CuSe. The k-points sampling is 1×1×1. For all the calculations, the bottom two layers of substrate atoms are fixed, while all the other atoms are fully relaxed. The structures were relaxed until the energy and residual force on each atom were smaller than $10^{-4}$ eV and 0.02 eV/Å, respectively. The vacuum layers of the two models are larger than 15 Å. The STM simulations are performed using the Tersoff-Hamann approach [30].

**Data availability.** The data that support the findings of this study are available from the corresponding author upon reasonable request.

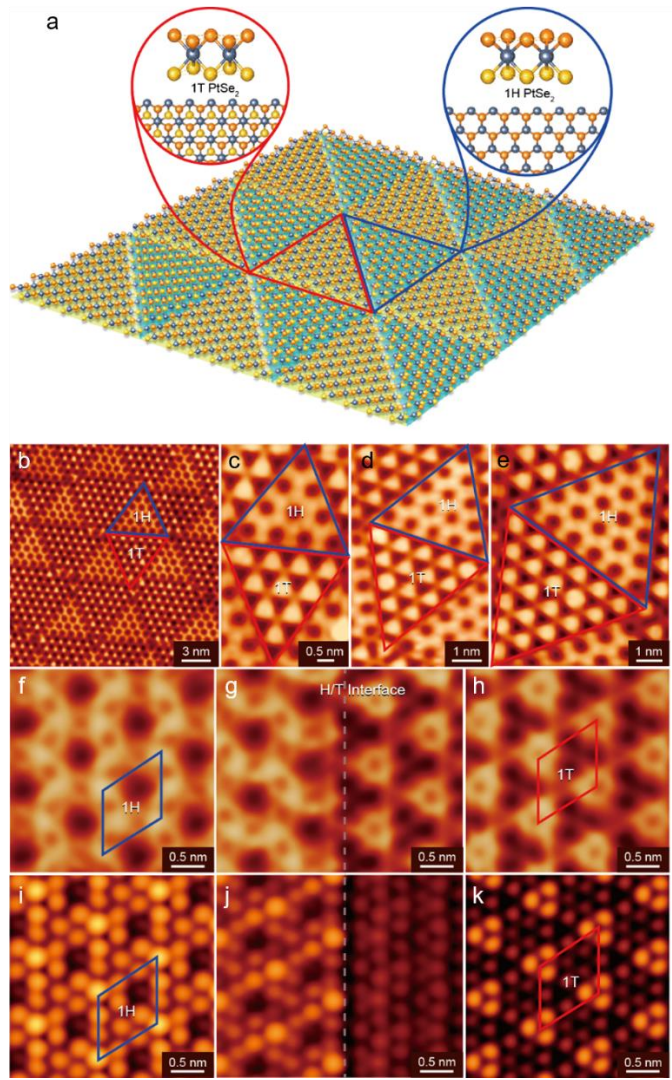

**Figure 1 | 1H/1T tiling pattern in monolayer PtSe₂. a**, Schematic illustration of the triangular tiling pattern formed by alternating 1H- and 1T-PtSe$_2$ areas, which are colored blue and yellow, respectively. Two views of atomic models for the two configurations are shown in the respective inserts. The top-layer Se atoms, Pt atoms and the bottom-layer Se atoms are colored orange, blue and yellow, respectively. **b**, STM image of the 1H/1T patterned structure in monolayer PtSe$_2$. $V_s$ = -1.6 V, $I_t$ = 0.1 nA. **c-e**, STM images of 1H- and 1T-PtSe$_2$ domains of different sizes. $V_s$ = -1.6 V, $I_t$ = 0.5 nA. **f-h**, High-resolution STM images of 1H-PtSe$_2$ (**f**), 1H/1T interface (**g**), and 1T-PtSe$_2$ (**h**). The unit cells of the 1H and 1T domains are marked by blue and red diamonds, respectively. $V_s$ = -1.0 V, $I_t$ = 0.8 nA. **i-k**, STM simulations of a 1H-PtSe$_2$ domain (**i**), the 1H/1T interface (**j**), and a 1T-PtSe$_2$ domain (**k**), respectively.

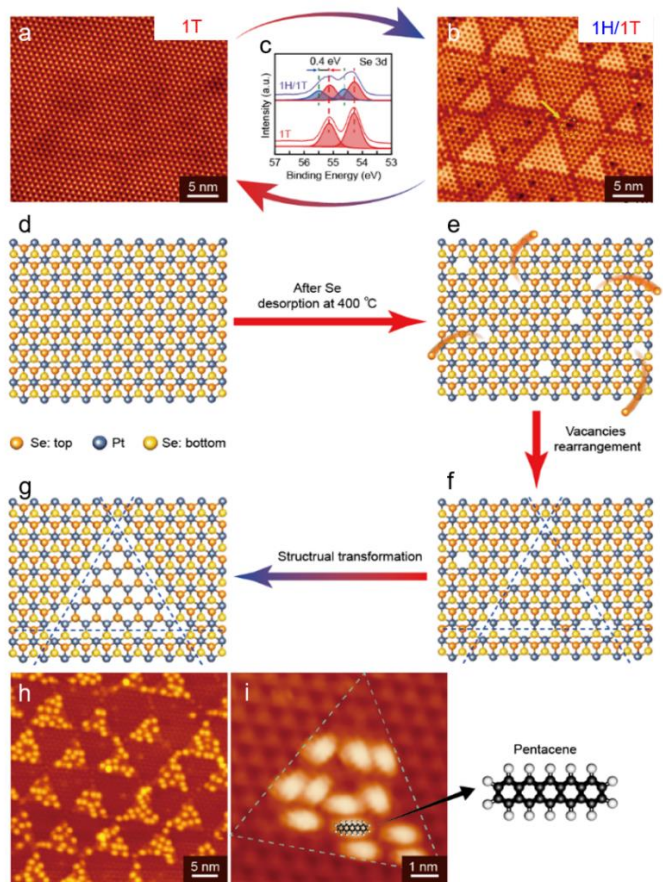

**Figure 2 | Reversible transition and selective adsorption of 1H/1T tiling pattern. a** and **b**, Typical STM images of pure 1T-PtSe$_2$ (**a**) and 1H/1T patterned PtSe$_2$ (**b**). The 1H/1T tiling structure can be obtained from a pure 1T structure by annealing the sample to 400°C. The reverse process (from 1H/1T tiling to pure 1T structure) can be realized by adding Se and annealing at 270°C. $V_s$ = -1.6 V, $I_t$ = 0.1 nA. **c**, The XPS spectra of Se 3$d$ electrons for the entire reversible structural transformation process. The red and purple peaks originate from Se 3d electrons in 1T- and 1H-PtSe$_2$ films with a 0.4-eV shift, respectively. **d-g**, Schematic illustration of the structural transformation from pure 1T-PtSe$_2$ to 1H/1T patterned PtSe$_2$. The formation of 1H-PtSe$_2$ domains is a result of the collective evolution of Se vacancies. **h**, An STM image of pentacene molecules selectively adsorbed on the 1H-PtSe$_2$ domains of a 1H/1T patterned PtSe$_2$ film, forming well-ordered triangular clusters. $V_s$ = -1.6 V, $I_t$ = 0.5 nA. **i**, High-resolution STM image of pentacene molecules selectively adsorbed on one 1H-PtSe$_2$ domain. A schematic of a pentacene molecule is shown on the right. $V_s$ = -1.0 V, I = 0.8 nA.

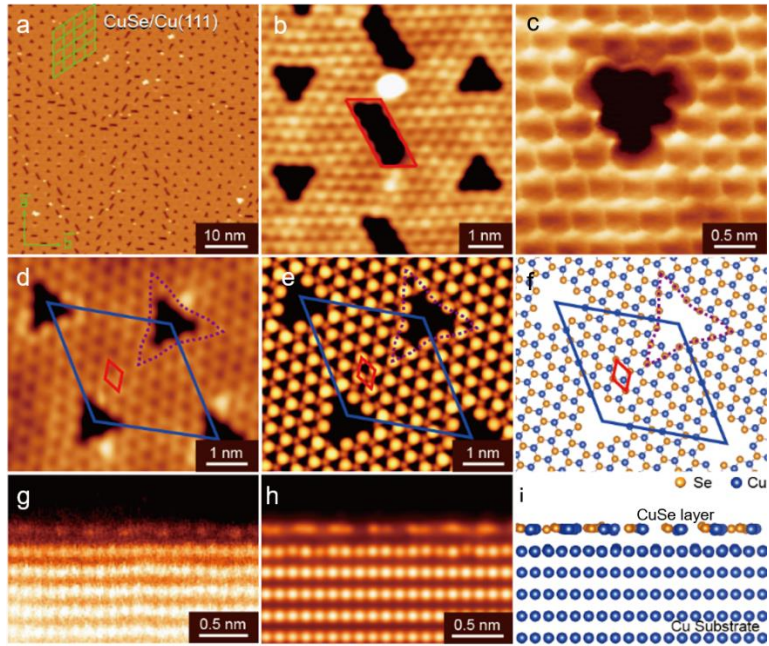

**Figure 3 | Patterned CuSe monolayer with periodic nanopores. a**, A large-scale STM image of patterned CuSe monolayer. The green lattice presents the 3-nm periodicity of the nanopores. $V_s = -1.5$ V, $I_t = 0.1$ nA. **b**, A high-resolution STM image of a domain boundary region. The boundary consists of parallelogram-shaped nanopores as indicated by the red parallelogram. $V_s = -1.0$ V, $I_t = 0.3$ nA. **c**, A high-resolution STM image of a single nanopore. Honeycomb lattice of CuSe and the zigzag edges of the nanopore are clearly resolved. $V_s = -0.01$ V, $I_t = 7.0$ nA. **d-f** An atomic-resolution STM image (**d**), simulated STM image (**e**), and theoretical model (**f**) of CuSe monolayer on Cu(111) with periodic nanopores, respectively. The images show the detailed structure of nanopores. $V_s = -1.0$ V, $I_t = 0.4$ nA. In **f**, Cu substrate is hidden for visualization purposes. **g-i**, Cross-section STEM image (**g**), simulated cross-section STEM image (**h**), and side view (**i**) of a CuSe monolayer on a Cu(111) substrate. The STEM image confirms that the patterned CuSe is single layer.

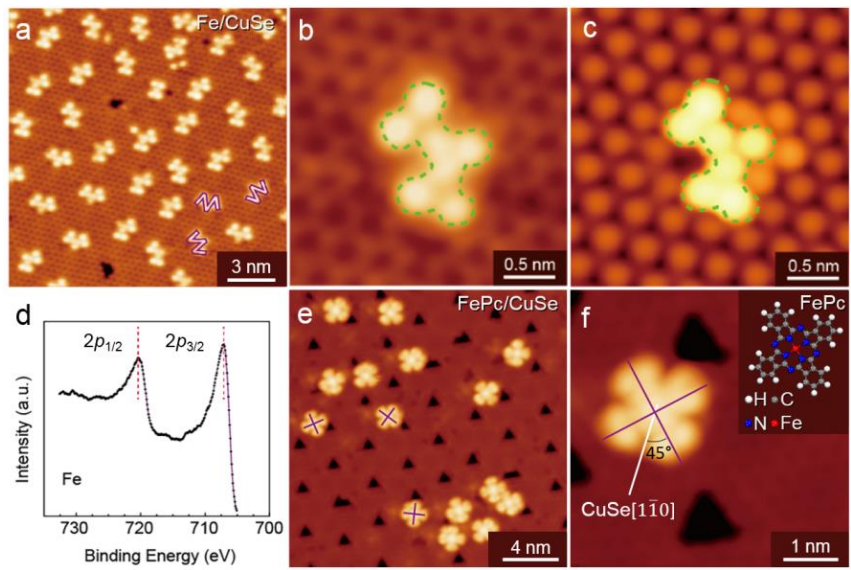

**Figure 4 | Selective adsorption on patterned CuSe with periodic nanopores. a**, A large-scale STM image of CuSe monolayer with W-shaped FeSe clusters. The clusters selectively adsorb in the nanopores of CuSe monolayer. $V_s$ = -0.1 V, $I_t$ = 1.5 nA. **b**, A high-resolution STM image of the W-shaped cluster. The image shows that the W-shaped cluster contains seven protrusions. $V_s$ = -0.1 V, $I_t$ = 1.5 nA. **c**, A simulated STM image of an $Fe_{13}Se_7$ cluster in the nanopores. The green contour line and the seven protrusions perfectly match those in (**b**). **d**, XPS data showing the Fe $2p_{3/2}$ and Fe $2p_{1/2}$ core levels. **e**, A large-scale STM image of FePc molecules on CuSe. The FePc molecules selectively adsorb on CuSe terraces with preferential orientations as discussed in the text. $V_s$ = -1.1 V, $I_t$ = 0.1 nA. **f**, High-resolution STM image of an FePc molecule on CuSe. Atomic structure of the FePc molecule is shown in inset. $V_s$ = -1.1 V, $I_t$ = 1.5 nA.

**Supplementary Text and Figures:**

**1. Formation of the moiré pattern.**

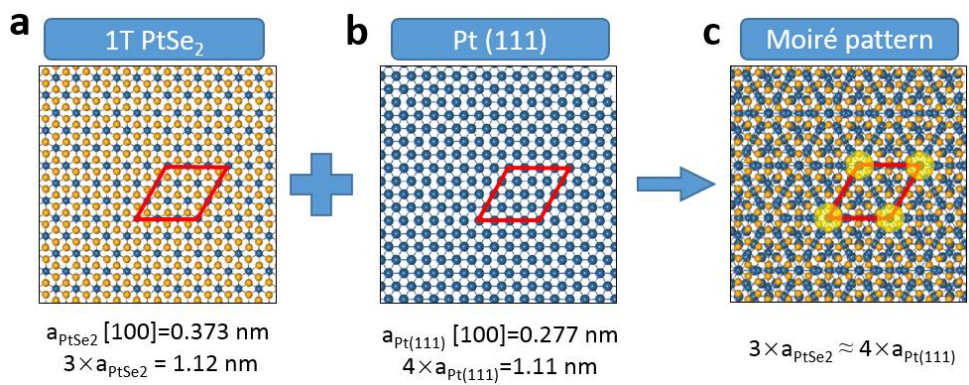

**Figure S1 | Models for the origin of the moiré pattern in PtSe$_2$ on Pt(111). a**, A model of the 1T PtSe$_2$ lattice. A (3×3) cell is marked by the red rhombus. **b**, A model of the Pt(111) lattice. A (4×4) cell is marked by the red rhombus. **c**, Because the size of a (3×3) PtSe$_2$ approximately equals to that of a (4×4) Pt(111), when a PtSe$_2$ monolayer sits on a Pt(111) substrate, moiré pattern forms with a (3×3)PtSe$_2$-on-(4×4)Pt(111) supercell.

## 2. The STM images, relaxed models and similated STM images of 1H- and 1T-PtSe$_2$ on Pt (111).

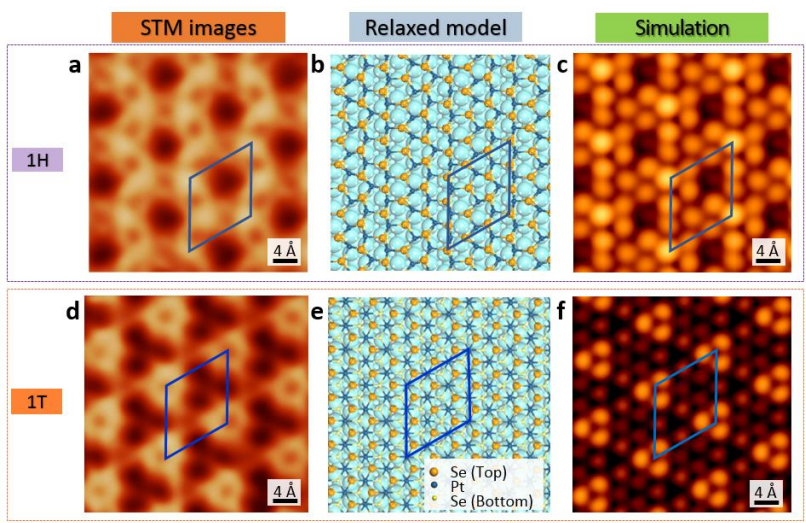

**Figure S2 | STM images, relaxed models and simulated STM images of 1H- and 1T-PtSe$_2$ on Pt (111). a**, STM image for 1H-PtSe$_2$ on Pt(111). $V_s$ = -1.0 V, $I_t$ = 0.8 nA. A unit cell of this moiré pattern is marked by the blue diamond. The brightest protrusions at the vertex of diamond are interpreted to be the Se atoms at atop site of Pt (111) surface. **b**, Top view of the relaxed model for 1H-PtSe$_2$ on Pt (111). **c**, STM simulation for 1H-PtSe$_2$ on Pt (111). The overall features of the experimental STM image (Fig. S2a) are well reproduced. **d**, STM image for 1T-PtSe$_2$ on Pt(111). $V_s$ = -1.0 V, $I_t$ = 0.8 nA. **e**, Top view of the relaxed model for 1T-PtSe$_2$ on Pt(111). **f**, STM simulation for 1T-PtSe$_2$ on Pt(111), which is also in good agreement with the STM image (Fig. S2d).

## 3. The tip effect in the STM image simulations of 1H-PtSe$_2$ on Pt(111).

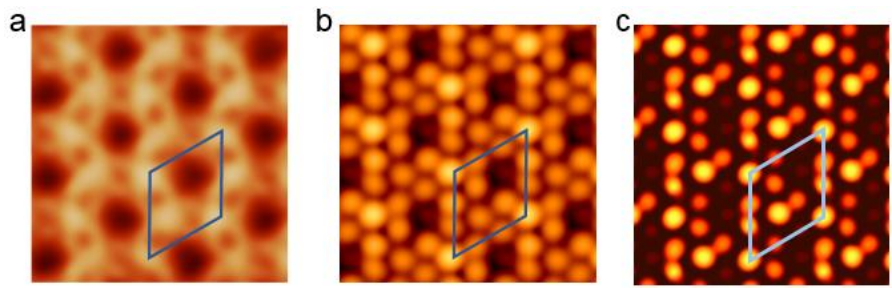

**Figure S3 | STM image simulations of 1H-PtSe$_2$ on Pt(111) with and without the tip effect. a**, A zoomed-in STM image. **b**, A simulated STM image without considering the tip effect. **c**, A simulated STM image considering the tip effect. As the simplest approximation, we model the tip by a single Se atom, corresponding to the case that a Se atom gets attached to the STM tip. With the tip effect, the spot inside the unit cell has similar brightness as the spots at the corners, in agreement with the data.

**4 The phonon dispersion of a free-standing 1H-PtSe$_2$ monolayer.**

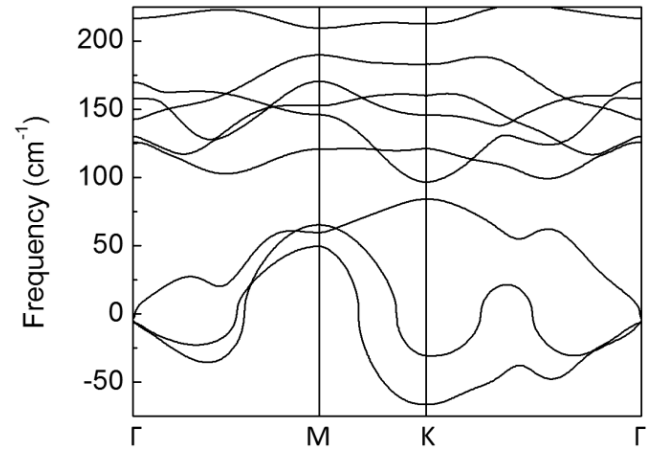

**Figure S4 | Phonon dispersion of a free-standing 1H-PtSe$_2$ monolayer.** There are negative-frequency phonon modes indicating the structure is not stable.

**5 Work function difference between 1H- and 1T- domains.**

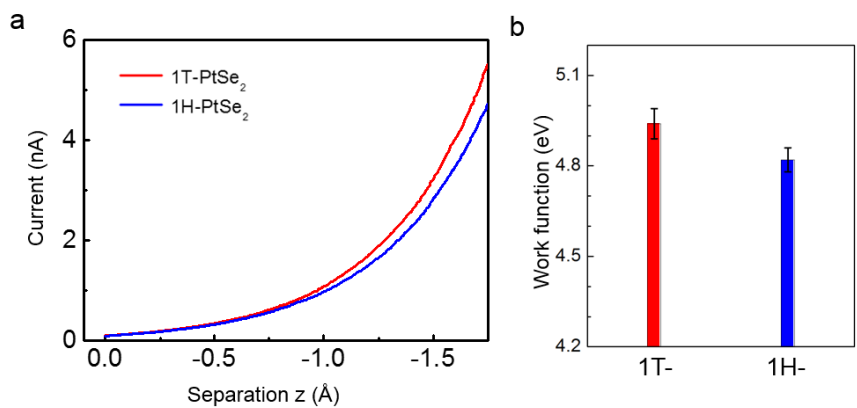

**Figure S5 | I(z) spectra and the derived local work function in 1T- and 1H- domains in a patterned 1H/1T-PtSe$_2$ film. a**, Averaged I(z) spectra and **b**, Derived local work function in 1T- and 1H- domains. The work function is derived with the method described in Ref. 31. In 1T-domains, the average local work function is 4.94 ± 0.05 eV. In 1H domains, the average local work function is 4.82 ± 0.04 eV. DFT calculations also reveal a larger work function for 1T-PtSe$_2$, 5.21 eV, compared with 5.15 eV for 1H-PtSe$_2$. These results provide further evidence of the existence of 1H-PtSe$_2$.

## 6. Atomic structures of 1T-PtSe$_2$ monolayer with 12.5% Se vacancies.

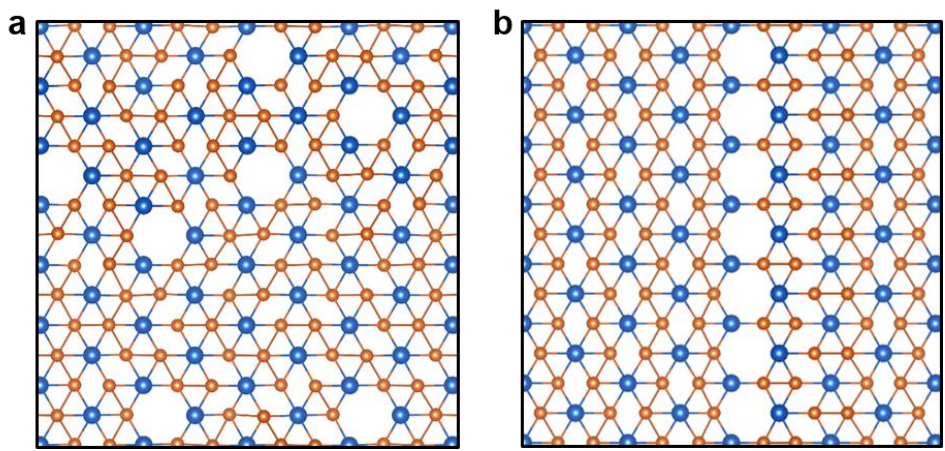

**Figure S6 | Atomic structures of 1T-PtSe$_2$ monolayer with 12.5% Se vacancies. a**, Se vacancies are random. **b**, Se vacancies form lines. DFT calculations indicate that vacancies in lines are more energetically favored by 0.02 eV per Se vacancy.

## 7. Atomic structures of Se vacancies surrounding PtSe₂ triangles.

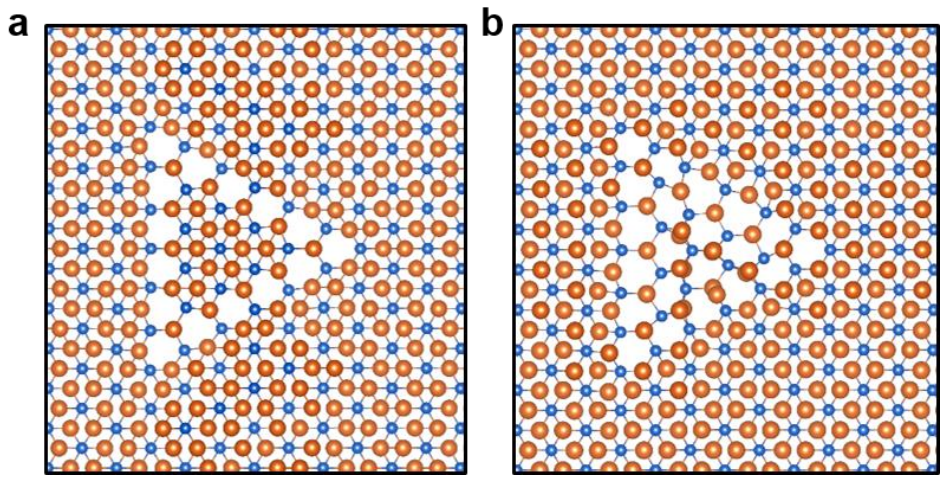

**Figure S7 | Atomic structures of Se vacancies surrounding PtSe₂ triangles. a**, PtSe₂ triangle in 1T phase. **b**, PtSe₂ triangle in 1H phase. DFT calculations indicate that PtSe₂ triangle in 1T phase is more energetically favored by 0.35 eV per Se atom (0.70 eV per PtSe₂ unit). It is worth noticing that in the presence of Se vacancy lines, the energy difference between 1H- and 1T-PtSe₂ is significantly lowered. For a perfect PtSe₂ monolayer, the energy difference between 1H- and 1T-PtSe₂ is calculated to be 1.33 eV per formula unit, while it is only 0.70 eV per formula unit in the presence of Se vacancy lines. The activation energy for the observed transformation should then be ~0.7 eV for each formula unit because the 1H phase is unstable. It is generally the case that activated processes, e.g., diffusion, occur at room temperature if the activation energy is ≤ 1 eV. For example, Ir atoms diffuse on Ir surfaces at room temperature with a migration barrier of 0.7 eV [32, 33]. Thus, the 1T-to-1H transformation that we observe can indeed occur after annealing at 400 °C. Similar phenomena have been observed in other TMD materials, where thermal annealing at elevated temperature (400 °C to 700 °C) leads to the formation of triangular inversion domains in MoS₂ [34] and 2H-to-1T phase transition in Re-doped MoS₂ [21].

# 8. Produce 1H/1T PtSe$_2$ triangular pattern with an off-center crucible to control the Se initial density.

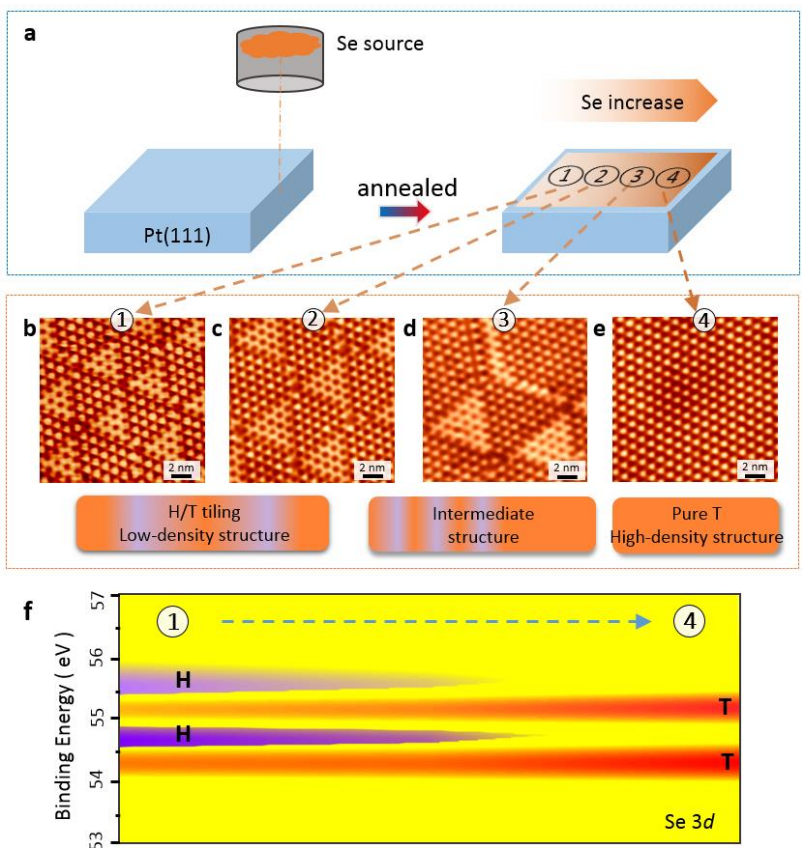

**Figure S8 | Produce 1H/1T triangular pattern with an off-center crucible to control the Se initial density. a**, Schematic diagram of the experimental method. The crucible filled with Se source is set at the edge instead of the center of substrate. The off-centered deposition leads to a varying initial density of Se atoms at the different positions of the sample. Four different areas with increasing Se density were chosen to investigate the structural variability of PtSe$_2$ films. **b**, STM image of a triangular tiling patterns of 1H and 1T PtSe$_2$ exist in the area ① with the relatively low initial Se density. **c, d**, The number of 1H-PtSe$_2$ domains decreases as the density of Se atoms increases, demonstrating intermediate structures (② ③). **e**, The continuously increasing Se density finally leads to 1T-PtSe$_2$ film (④). **f**, The XPS mapping of Se 3$d$ electrons along the direction of increasing Se density areas (area ① to area ④) demonstrates a significant decrease of the 1H-PtSe$_2$ and increase of the 1T-PtSe$_2$. For all the STM images, $V_s$ = -1.0 V, $I_t$ = 0.1 nA.

## 9. *in situ* LEED and XPS characterizations of 2D patterned CuSe film.

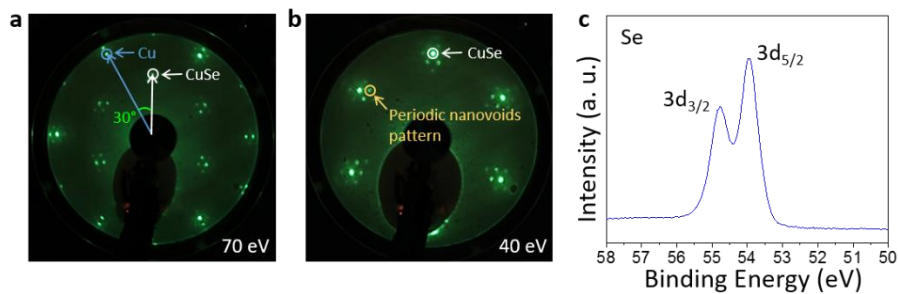

**Figure S9 | LEED and XPS characterizations of 2D patterned CuSe film. a**, **b**, Two LEED patterns obtained in the full layer patterned CuSe sample on the Cu(111) substrate at 70 eV (**a**) and 40 eV (**b**), respectively. The outer six diffraction spots in (**a**) (marked by blue circle) are assigned to the six-fold symmetry of the Cu substrate and the inner six diffraction spots (marked by white circle) are assigned to the three-fold symmetry of the CuSe lattice. The figure clearly shows that the angle between the CuSe lattice and the Cu substrate is 30°, which is in good agreement with the STM data. In addition, the diffraction spots (marked by yellow circle in (**b**)) surrounding the CuSe diffraction spots are induced by well-defined hexagonal periodic nanopores in the 2D patterned CuSe film observed in the STM images. **c**, The core level XPS spectrum of patterned CuSe sample on Cu(111) substrate. Two perfect singlet peaks can be resolved in the Se $3d$ signal, 54.77 eV and 53.95 eV, which correspond to the binding energy positions of Se $3d_{3/2}$ and $3d_{5/2}$ peaks, respectively, indicating the dominance of $Se^{2-}$ chemical states in the CuSe sample.

## 10. Atomic models to explain the formation of 13-atoms-missing pore.

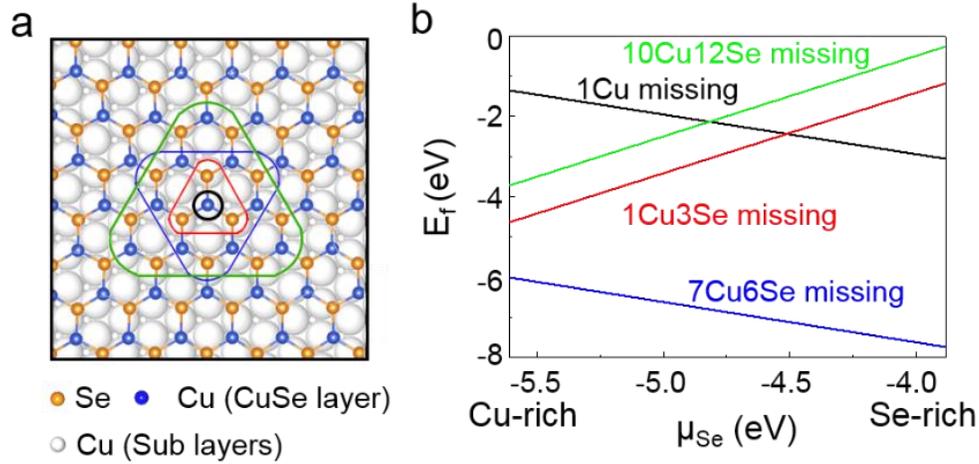

**Figure S10 | Formation energies of pores with different sizes in CuSe/Cu(111). a**, Atomic configurations to show pores with different possible size. Black, red, blue, and green circles correspond to 1Cu, 1Cu3Se, 7Cu6Se, and 10Cu12Se atoms missing, respectively. **b**, Formation energy of pores with different sizes. The pore with 13 atoms missing is always energetically favored. Formation energies are negative, corresponding to energy gain. It is clear that the formation of 7Cu6Se pores is energetically preferred and occurs spontaneously.

The relative stability of nanopores with different size was tested by comparing their formation energies. The formation energy is a function of the chemical potential of Se:

$$E_f = E_{w/\ pore} + N_{Cu}\ [E_{CuSe} - \mu_{Se}] + N_{Se}\mu_{Se} - E_{w/o\ pore}$$

where $E_{w/\ pore}$ and $E_{w/o\ pore}$ are the total energies with and without pores. $N_{Cu}$ is the number of missing Cu atoms. $N_{Se}$ is the number of missing Se atoms. $\mu_{Se}$ is the chemical potential of Se. For Se-rich conditions, $\mu_{Se}$, was chosen as the energy of a Se atom in $Se_8$ molecules. For Se-poor condition, $\mu_{Se}$ is chosen as $\mu_{Se} = E_{CuSe} - E_{Cu\text{-bulk}}$, which reflects plentiful Cu atoms in bulk form.

## 11. Comparison between 13 Fe and 15 Fe atoms in a nanopore in CuSe monolayer.

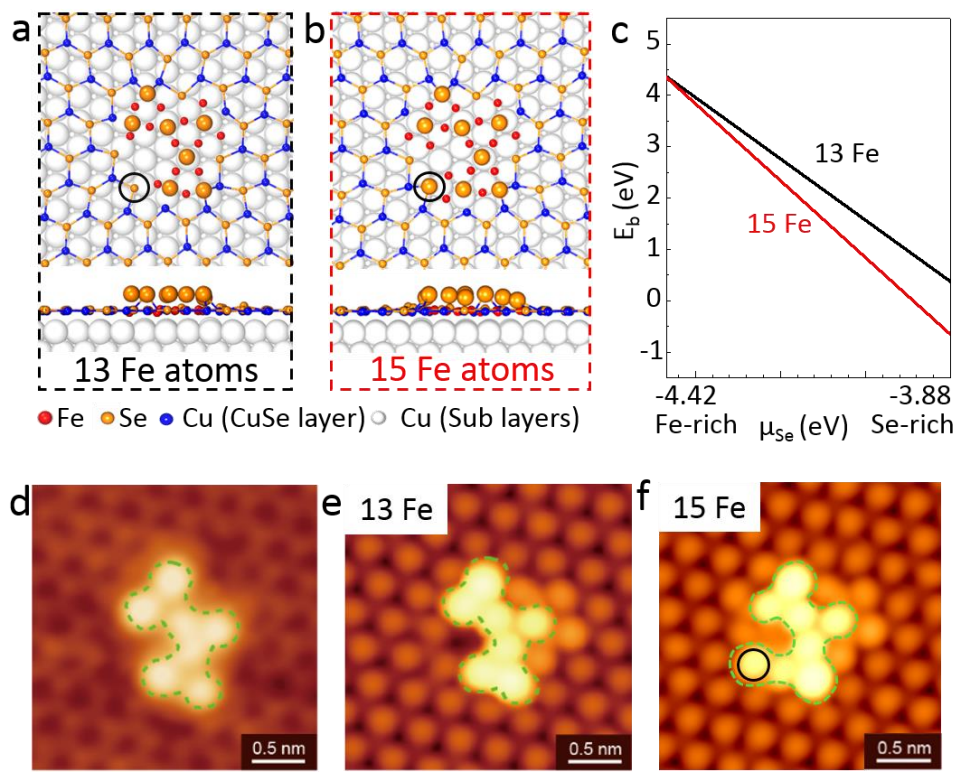

**Figure S11 | Comparison between 13 Fe and 15 Fe atoms in a nanopore in CuSe monolayer. a, b,** Top views and side views of atomic models of $Fe_xSe_y$ clusters with 13 (**a**) and 15 Fe atoms (**b**), respectively. In the 15 atom-configuration, there is one more elevated Se atom (labeled by black circles). **c,** Binding energy of clusters with 13 and 15 Fe atoms as function of $\mu_{Se}$, the chemical potential of Se. $\mu_{Se}$ is chosen in the range of -4.42 eV < $\mu_{Se}$ < -3.88 eV, corresponding to Fe-rich and Se-rich cases, respectively. For virtually all values of $\mu_{Se}$, clusters with 13 Fe atoms are energetically favored. **d,** Experimental STM image. **e,** Simulated image with 13 Fe atoms. **f,** Simulated image with 15 Fe atoms. There is an extra bright spot for the 15 Fe atoms configuration (labeled by black circle). Note that the elevated (bright) atoms are Se. It is clear that the model with 13 Fe atoms in a pore is in better agreement with the experimental data.

## 12. Fe clusters in parallelogram-shape nanopores at domain boundary.

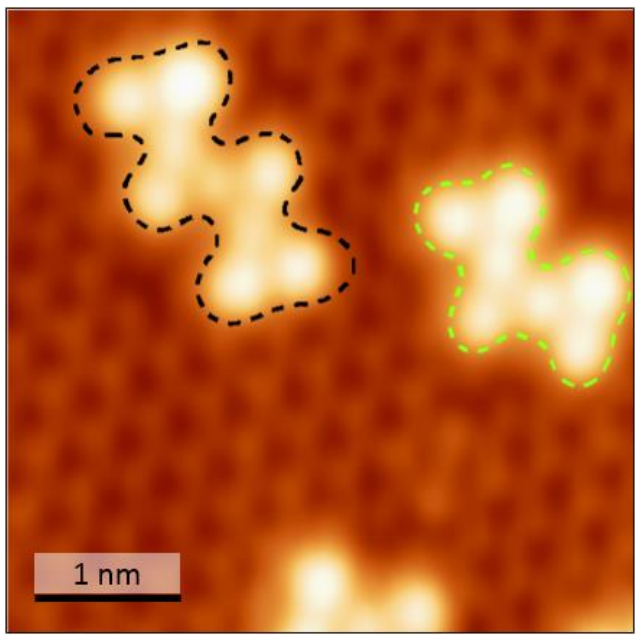

**Figure S12 | A high-resolution STM image showing a zigzag-shaped cluster in a parallelogram-shape nanopore at domain boundary (indicated by the black dashed line).** For comparison, a W-shaped cluster in a triangular nanopore is also shown (indicated by the green dashed line). The zigzag-shaped cluster can be viewed as an extended W-shaped cluster. $V_s$ = -0.1 V, $I_t$ = 1.5 nA.

## 13. Stability tests in ambient conditions.

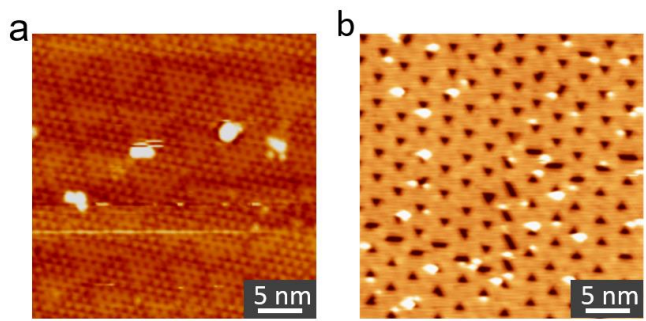

**Figure S13 | STM images of intrinsically patterned materials to prove the stability in ambient conditions. a**, 1H/1T- PtSe$_2$ on a Pt(111) substrate. $V_s$ = -1.6 V, $I_t$ = 0.1 nA. **b**, CuSe on a Cu(111) substrate. $V_s$ = -1.5 V, $I_t$ = 0.1 nA. The samples were exposed in air for more than 12 hours, then put back in the high-vacuum chamber and annealed at 200 °C for PtSe$_2$ and 400 °C for CuSe to remove possible adsorbates. The STM images show that, except for some contaminants (big bright spots), the samples keep their patterns.

## 14. Adsorption of different molecules on 1H and 1T PtSe$_2$ on Pt(111).

**Table S1 | Binding energies of different molecules on 1H- and 1T-PtSe$_2$ on Pt(111) in a unit of eV.** Calculation results show that these molecules prefer to adsorb on the 1H region.

|  | on 1H-PtSe$_2$/Pt(111) | on 1T-PtSe$_2$/Pt(111) |
| --- | --- | --- |
| NO$_2$ | 1.28 | 0.38 |
| CO | 0.32 | 0.02 |
| CO$_2$ | 0.20 | 0.05 |
| C$_2$H$_4$ | 0.46 | 0.18 |
| C$_{22}$H$_{14}$ | 1.31 | 1.00 |